\pdfoutput=1
\documentclass[lettersize,journal]{IEEEtran}
\usepackage{amsmath,amsfonts}
\usepackage{array}
\usepackage{textcomp}
\usepackage{stfloats}
\usepackage{url}
\usepackage{verbatim}
\usepackage{graphicx}
\makeatletter
\let\MYcaption\@makecaption
\makeatother
\usepackage[font=footnotesize]{subcaption}

\makeatletter
\let\@makecaption\MYcaption
\makeatother
\usepackage{cite}
\graphicspath{{./figures/}}
\usepackage{amsthm}
\theoremstyle{remark}            \newtheorem{remark}{Remark}      

\usepackage{amssymb}
\usepackage{algorithm}
\usepackage{algpseudocode}
\floatname{algorithm}{Protocol}
\usepackage{float}
\usepackage{threeparttable}
\usepackage{color}
\usepackage{tabularx}
\usepackage{booktabs} \usepackage{makecell}
\usepackage{ragged2e} \usepackage[hidelinks]{hyperref}

\begin{document}
    
    \title{Filter-Augmented Reinforcement Learning for Battery Fast Charging with State of Health and Plating-Aware Reward Design}
    
    \author{Meng Yuan,~\IEEEmembership{Senior Member,~IEEE}, Yang Li,~\IEEEmembership{Senior Member,~IEEE}, and Changfu Zou,~\IEEEmembership{Senior Member,~IEEE}
        \thanks{This work was supported by the European Union's Horizon Europe programme under the Marie Skłodowska-Curie Actions Postdoctoral Fellowships (Grant No.~101110832), and the Swedish Research Council (Grant No.~2023-04314). The mobility of C.~Zou was funded by the Swedish Foundation for International Cooperation in Research and Higher Education (STINT; Grant No.~MG2020-9107).}\thanks{Meng Yuan is with the School of Engineering and Computer Science, Victoria University of Wellington, Kelburn 6012, New Zealand (email: {meng.yuan@vuw.ac.nz})}
        \thanks{Changfu Zou is with the Department of Electrical Engineering, Chalmers University of Technology, 412 96 Gothenburg, Sweden. (email: {changfu.zou@chalmers.se})}
        \thanks{Yang Li is with the School of Electrical Engineering and Automation, Wuhan University, Wuhan 430072, China. (email: {yangli@ieee.org})}
    }

    \markboth{Preprint}{Filter-Augmented Reinforcement Learning for Battery Fast Charging with State of Health and Plating-Aware Reward Design}

    \maketitle
    
    \begin{abstract}
        Fast charging of lithium-ion batteries is crucial for electric vehicle adoption, yet it increases safety risks and accelerates battery ageing. To address this, we propose a filter-augmented reinforcement learning (RL) framework for lifelong health-aware fast charging, based on the twin-delayed deep deterministic policy gradient (TD3) algorithm. The RL agent is guided by a reward function that balances charging speed and battery longevity through the incorporation of state of health (SoH) information and an SoH-dependent voltage penalty term derived from a constant--current constant--overpotential charging strategy. Since onboard SoH estimates are inevitably corrupted by noise, two filtering strategies are integrated: an exponential moving average (EMA) filter when noise statistics are unknown, and a Kalman filter when they are available, with the filtered estimate embedded in both the agent's state and reward computation. The RL policy is trained using data from high-fidelity life-cycle simulations. The evaluation uses a published electrochemical and degradation parameterisation for a commercial LG M50 cell. Results show that the proposed method achieves up to 21.4\% more equivalent full cycles (EFCs) than conventional constant current--constant voltage (CC--CV) charging under 2\% SoH estimation noise, and maintains this lifetime advantage even under 5\% noise, while sustaining comparable charging speed.
    \end{abstract}

    
    \section{Introduction}
    
    \IEEEPARstart{D}{riven} by the global push towards electrification and sustainable energy, lithium-ion batteries (LIBs) have become the dominant energy storage solution across a wide range of applications, including consumer electronics, electric vehicles (EVs), and large-scale renewable energy systems \cite{zubi2018lithium, bobobee2025state}. Despite ongoing advances in battery chemistry, battery management systems (BMS) have become essential for achieving higher performance from existing LIB technologies by implementing charging strategies within safe material limits and operating boundaries \cite{attia2020closed}.
    
    A core function of the BMS is to reduce charging time while preserving long-term battery health \cite{attia2020closed}. Achieving this goal is challenging, as aggressive charging protocols often exacerbate degradation mechanisms such as lithium plating, solid electrolyte interphase (SEI) growth, and thermal instability \cite{koleti2021development}. Among these, lithium plating is particularly prevalent under high charging currents and low temperatures, and it can exponentially accelerate capacity fade and poses serious safety risks \cite{yang2018understanding}.
    The competing priorities of high charging speed and long battery lifetime highlight the need for advanced control strategies that can intelligently regulate charging behaviour in a health-aware manner. Such approaches must account for the complex interactions between external stress factors and battery internal dynamics, thereby enabling safer and more efficient utilisation of LIBs \cite{tomaszewska2019lithium}.
    
    Despite the significant advancements in BMS capabilities, today's LIBs are still charged using conventional model-free methods such as constant current--constant voltage (CC-CV) and its variants. While simple to implement and widely adopted, CC-CV charging can only account for battery degradation in a very coarse, heuristic way, often resulting in prolonged charging times and premature degradation.

    Beyond these conventional methods, a range of alternative fast-charging protocols has been developed, including multistage constant-current charging, pulse charging, feedback control, and optimisation-based current profiles. Pulse-charging performance depends on the pulse amplitude, duty cycle, frequency, cell design, and thermal condition, and is therefore not universally superior to a matched continuous-current protocol \cite{tomaszewska2019lithium,koleti2021development}.

    Efforts to achieve faster charging typically involve raising current rates or voltage thresholds within the CC-CV framework. However, such modifications often lead to accelerated degradation and elevated safety issues \cite{huang2022onboard,lin2021lithium}. To mitigate this issue, health-aware feedback charging strategies have emerged, which integrate degradation mechanisms or related indicators into closed-loop control. For example, in \cite{wassiliadis2023model}, a proportional-integral-derivative (PID) control strategy using anode overpotential as feedback was proposed to suppress lithium plating and improve charging efficiency. Following the trend, a reinforcement learning (RL)-based fast charging strategy was developed to enforce an anode overpotential constraint \cite{yang2022constant}. A central assumption underpinning these studies is that the anode overpotential is directly measurable during operation. Although anode overpotential is accessible via a reference electrode in laboratory settings, this invasive setup is impractical for commercial deployment.
    
    Model-based approaches have also been widely used for fast battery charging to bypass the need for additional internal sensors. Electrochemical-model-based strategies incorporating state estimation and PI control \cite{liyang2021tec}, as well as nonlinear model inversion \cite{liyang2023tcst_inversion}, offer computationally efficient alternatives to model predictive control (MPC). However, these results were obtained over one or a few charging cycles without accounting for long-term degradation. To address this lifespan limitation, recent works have employed MPC frameworks with lifelong parameter updates \cite{qasem2025real, qasem2025dynamic, lu2024health}. These methods still keep an online model and solve an optimisation problem during operation, but how demanding this is depends on the model order, solver, and hardware.

    One such recent example is an AI-enabled digital-twin framework that estimates capacity and internal resistance from partial-discharge voltage, current, and temperature data. These estimates update an electrochemical ageing model used by online MPC to determine the charging current. The study also tests the health estimator under injected voltage-sensor noise \cite{qasem2026digital}. In this framework, learning supports health estimation rather than directly implementing the charging controller.

    Representative health-aware charging methods, including degradation-aware MPC and adaptive RL \cite{hwang2022model,hao2023adaptive}, differ in the health indicators used, the required internal-state estimation, and the online model and optimisation burden, yet none has been evaluated over the full battery lifetime under noisy SoH feedback.
    
    To resolve this computational bottleneck while handling complex safety boundaries, data-driven methods such as RL offer a promising alternative by shifting the optimisation process to an offline training phase. Existing RL fast charging strategies have successfully incorporated temperature and multiphysics constraints \cite{wei2021deep, park2022deep}. Nevertheless, similar to the early model-based approaches, these RL strategies predominantly focus on single-cycle optimisation for a battery with fixed parameters. If the static policies trained in these works are applied to an aged battery, the increased internal resistance and shifted electrochemical boundaries could lead to severe lithium plating. Furthermore, while some recent RL-based strategies have begun to consider battery degradation over the lifespan \cite{wang2026health}, they primarily focus on macroscopic scheduling and energy cost minimisation rather than addressing the microscopic physical constraints strictly required for safe fast charging. Therefore, a critical need remains for a fast charging strategy that retains the computational efficiency of RL while enabling lifelong adaptation to battery degradation and strict adherence to instantaneous physical safety boundaries.
    
    In this work, SoH is capacity-based and is defined as the ratio of the presently available capacity to the nominal capacity of the fresh cell. Such lifelong adaptation relies on continuous SoH feedback from the BMS. However, in practice, onboard SoH estimates are inevitably corrupted by noise. Directly using such noisy SoH values in an RL-based charging controller introduces two problems. First, the SoH-dependent voltage mapping used to approximate lithium plating boundaries becomes inaccurate. Second, the RL policy itself may exhibit erratic behaviour and converge to sub-optimal solutions due to stochastic perturbations in its state input. Despite practical importance, the effect of SoH estimation noise on control performance has been largely neglected in existing RL-based charging studies.

    To address the research gaps identified above, this work proposes a twin-delayed deep deterministic policy gradient (TD3)-based RL framework for lifelong health-aware fast charging. The reward function incorporates state of charge (SoC), SoH, and an ageing-dependent cut-off voltage derived from a constant current--constant overpotential (CC--COP) protocol as a proxy for lithium plating risk. To handle the inevitable noise in onboard SoH estimates, two filters are integrated depending on available prior knowledge: an EMA filter when noise statistics are unknown, and a Kalman filter (KF) when they are known. The filtered SoH replaces the raw measurement in both the agent's state and reward computation. The framework relies only on SoC, SoH, and terminal voltage, all of which are available in commercial BMSs, and requires no physical model at deployment. Unlike the model-based methods above, the TD3 actor directly generates the charging current. Online operation therefore requires actor inference and scalar-filter updates, but no electrochemical-model execution or repeated optimisation. Moreover, this work directly filters and evaluates 2\% and 5\% SoH-estimation noise at the controller interface.
    
    The rest of this paper is organised as follows. In Section~\ref{sec:problem_charging}, the health-aware fast charging problem for lithium-ion batteries is formulated, and the CC-COP charging strategy is described. Section~\ref{sec:method} presents the proposed RL-based charging approach, including the robust SoH processing via filtering. Section~\ref{sec:results} presents the battery modelling framework, experimental setup, and simulation results, including detailed analyses of ageing effects and lithium plating. Finally, conclusions are drawn in Section~\ref{sec:conclusion}.
    
    \section{Problem Formulation}\label{sec:problem_charging}
    
    In applications such as EVs, a battery is generally considered to have reached the end of its useful life when its SoH declines to a threshold, typically 75--80\% of its original capacity \cite{waldmann2018li,zhu2021end}. At the same time, users increasingly expect shorter charging times during each use. Therefore, an effective charging strategy must achieve fast charging while maintaining long-term battery health. This dual objective forms the foundation of health-aware fast charging. To meet this objective, we aim to minimise the time required to reach the target SoC in each cycle while maximising the number of EFCs before the SoH falls below the threshold.
    
    \subsection{Formulation of a General Lifelong Charging Optimisation Problem} \label{sec:2_1}
    
    To formulate a general lifelong charging optimisation problem, we first denote the charging starting and end time instants for the $i$th cycle as $t_0(i)$ and $t_f(i)$, respectively. The battery current can be expressed as
    \begin{equation} \label{eq:current}
        I(t) = u(t) = \begin{cases}
            -u_{\text{ch}}(t), & t \in [t_0(i), t_f(i)], \forall i \\
            u_{\text{ref}}(t),  &  \text{otherwise,}
        \end{cases}
    \end{equation}
    where $u_{\text{ch}}\geq0$ represents the charging-current magnitude to be optimised, so the signed battery current is negative during charging. The signal $u_{\text{ref}}$ denotes the fixed current during all other phases.

    The battery current $u$ and battery measurements $y$ are subjected to a dynamic model, given by
    \begin{align} \label{eq:bat_model}
        \begin{cases}
            \dot{x}(t) = f(x(t),u(t)), \\
            y(t) = h(x(t),u(t)),
        \end{cases}
    \end{align}
    where $x$ denotes the state vector, and the output $y$ involves measurable or estimated outputs such as SoC, SoH, voltage $V$, and anode overpotential $\eta_{\text{side}}$.
    
    Since battery lifespan is significantly affected by degradation mechanisms such as lithium plating, and low anode overpotentials are known to accelerate this process \cite{zhang2025machine}, a health-aware fast charging problem can be formulated as the minimisation of the following cost function:
    \begin{equation}
        J(t) = \lambda_{1} C_{\text{SoC}}(t) + \lambda_{2} C_{\text{side}}(t),
    \end{equation}
    where \(\lambda_{1}\) and \(\lambda_{2}\) are nonnegative weighting factors. The first term, $C_{\text{SoC}}(t)$, corresponds to SoC setpoint tracking and is designed to reduce charging duration. It is defined as:
    \begin{equation}
        C_{\text{SoC}}(t) = \left| \text{SoC}^* - \text{SoC}(t) \right|,
    \end{equation}
    where \(\text{SoC}^*\) denotes the user-defined target SoC upon completion of the charging process, and \(\text{SoC}(t)\) is the present SoC at time instant $t$.
    
    To reduce the degradation effects, particularly due to lithium plating, the second term can be defined as:
    \begin{equation}
        C_{\text{side}}(t) = \left| \eta_{\text{side}}(t) - \eta_{\min} \right|,
    \end{equation}
    where $\eta_{\min}$ is a predefined threshold for anode overpotential, below which the lithium plating is likely to occur.
    \begin{remark}
        Note that, unlike many existing studies that pursue plating-free strategies for battery chemistries suitable for fast charging in EVs, this work relaxes such constraints by not imposing strict limits on anode overpotential. This approach acknowledges that, for certain battery chemistries, some degree of lithium plating may be unavoidable in order to meet the required charging speed.
    \end{remark}

    The decision variable in this problem, namely the charging-current magnitude $u_{\text{ch}}$, is subject to physical constraints imposed by the charger:
    \begin{equation}
        0 \leq u_{\text{ch}}(t) \leq I_{\max}. \label{eq:I_constraint}
    \end{equation}
    
    Let $\text{SoH}_i(u_{\text{ch}})$ denote the SoH after cycle $i$. The policy-dependent end-of-life cycle index is
    \begin{equation}
        N_{\text{EOL}}(u_{\text{ch}})
        =\min\left\{i\in\{1,2,\ldots\}\mid
        \text{SoH}_i(u_{\text{ch}})\leq\text{SoH}_{\text{end}}\right\}.
        \label{eq:eol_cycle}
    \end{equation}
    Let $Q_i^{\text{ch}}(u_{\text{ch}})$ denote the total charged capacity in cycle $i$. With fresh-cell nominal capacity $Q_{\text{nom}}$, the cumulative EFCs at end of life are
    \begin{equation}
        N_{\text{EFC}}(u_{\text{ch}})
        =\frac{1}{Q_{\text{nom}}}
        \sum_{i=1}^{N_{\text{EOL}}(u_{\text{ch}})}
        Q_i^{\text{ch}}(u_{\text{ch}}).
        \label{eq:efc_definition}
    \end{equation}
    The health-aware fast charging problem is then formulated as the following bi-objective optimisation problem:
    \begin{align}
        \underset{u_{\text{ch}}(t)}{\min} & \quad
        \left[
        \frac{1}{N_{\text{EOL}}(u_{\text{ch}})}
        \sum_{i=1}^{N_{\text{EOL}}(u_{\text{ch}})}
        \int_{t_0(i)}^{t_f(i)}J(t)\,dt,\,
        -N_{\text{EFC}}(u_{\text{ch}})
        \right] \nonumber \\
        \text{s.t.} & \quad \text{Battery dynamics:} \quad \eqref{eq:bat_model}, \nonumber \\
        & \quad \text{Input constraints:} \quad \eqref{eq:current}, \eqref{eq:I_constraint},  \nonumber \\
        & \quad V(t)\leq V_{\text{hard}}, \nonumber \\
        & \quad \text{SoC}(t_{0}(i)) = \text{SoC}_{0}(i), \nonumber \\
        & \quad \text{SoC}(t_{f}(i)) = \text{SoC}^{*}, \nonumber \\
        & \quad i=1,\ldots,N_{\text{EOL}}(u_{\text{ch}}). \label{eq:problem_formulation}
    \end{align}
    The first objective averages the explicitly defined cost $J(t)$ over completed charging cycles, so a longer-lived policy is not penalised merely for operating over more cycles. The second objective maximises lifetime throughput in EFCs. The vector objective is interpreted in the Pareto sense. Here, $V_{\text{hard}}$ denotes the hard terminal-voltage limit. The initial SoC $\text{SoC}_0(i)$ may vary in the general formulation. The evaluated protocol uses $\text{SoC}_0(i)=20\%$, $\text{SoC}^{*}=80\%$, and $\text{SoH}_{\text{end}}=80\%$ for every cycle.

    Equation \eqref{eq:problem_formulation} defines the general lifelong charging problem and its competing objectives. It is not an online optimisation solved at each control instant. Section~\ref{sec:method} introduces TD3 as a learning-based approach to address this problem, not as an exact solver or explicit scalarisation of \eqref{eq:problem_formulation}. The competing requirements of charging speed, degradation mitigation, and battery lifetime are reflected in the scalar reward design. Through offline interaction with the lifelong battery environment, TD3 learns a state-dependent charging policy. During deployment, the trained actor directly generates the signed charging current from the observed state.
    
    \subsection{Constant Current--Constant Overpotential Based Control}\label{sec:CC-COP}
    
    To solve the fast charging problem as shown in \eqref{eq:problem_formulation}, an intuitive approach is to design the controller based on overpotential feedback. This CC-COP method mirrors the traditional CC-CV approach by beginning with a constant current phase, followed by dynamically regulating the anode overpotential to a small positive value using real-time feedback. This method has been shown to mitigate lithium plating by maintaining the anode overpotential within a safe range after the initial current phase \cite{yang2022constant,zhang2025machine}.
    
    Despite its effectiveness, CC-COP control faces two limitations for practical fast charging: First, direct measurement of anode overpotential is difficult in commercial batteries, as it typically requires specialised sensors available in laboratory setups. Second, selecting an appropriate overpotential threshold is non-trivial. A high threshold can reduce the chance of lithium plating but it can also result in overly conservative charging, leading to longer charging times. This sensitivity to threshold tuning complicates real-world implementation and limits the broader adoption of CC-COP strategies.

    \section{Reinforcement Learning-Based Fast Charging Algorithm}\label{sec:method}
    
    \begin{figure}[tb]
        \centering
        \includegraphics[width=0.95\linewidth,keepaspectratio]{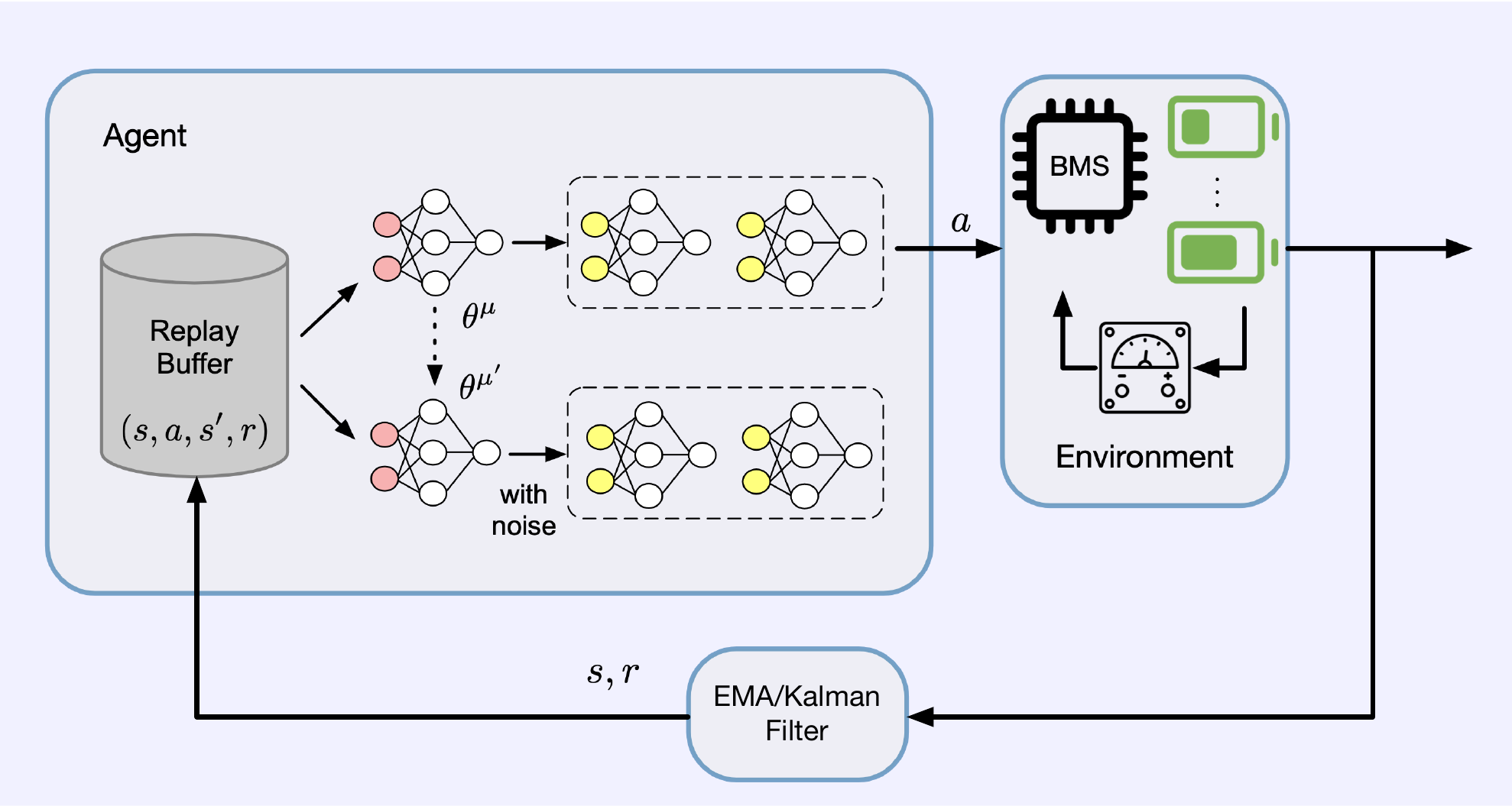}
        \caption{Structure of the proposed method.}
        \label{fig:entire_system2}
    \end{figure}
    
    To address the limitations of CC-COP control and the challenges associated with model-based charging strategies, we propose a data-driven RL approach for health-aware fast charging. Unlike traditional methods that rely on accurate physical modelling and parameter calibration, RL learns optimal control policies through direct interaction with the environment, guided by observable states and reward signals. In this work, we develop a TD3-based fast charging strategy that explicitly considers battery degradation caused by lithium plating, which is strongly influenced by anode overpotential. While direct measurement of overpotential is impractical in real-world applications, we adopt a CC-COP protocol in a laboratory setting to extract degradation-related information. Moreover, although many studies have explored the estimation of battery SoH \cite{fang2019online,shen2023accurate}, existing RL-based charging strategies \cite{wei2021deep,park2022deep,wang2026health} have not incorporated real-time SoH feedback into the charging control process. To address this gap, our proposed method integrates SoH information into both the training and execution of the RL agent, enabling adaptive, health-aware charging throughout the battery life-cycle. The overall system architecture is shown in Fig.~\ref{fig:entire_system2}.
    
    \subsection{SoH-Voltage Correlation for Health-Aware Charging}
    
    \begin{figure*}[t]
        \centering
        \includegraphics[width=0.6\textwidth,keepaspectratio]{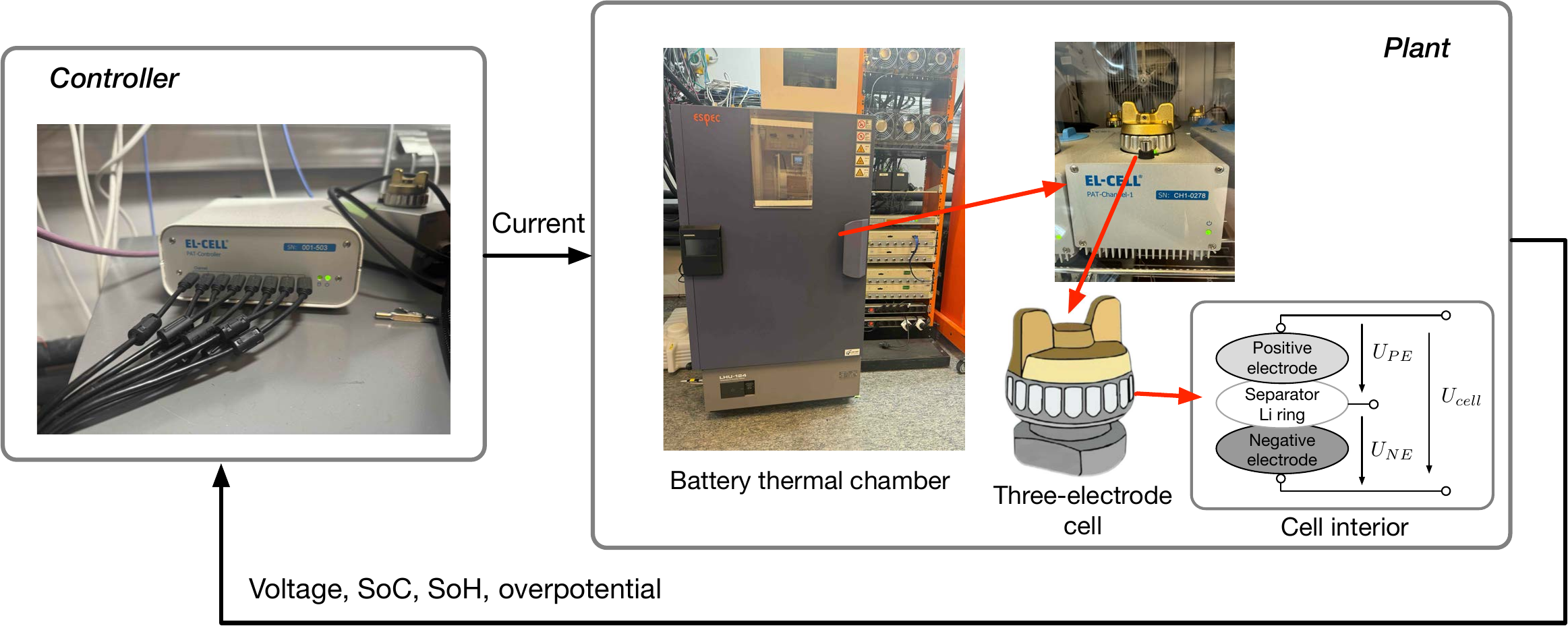}
        \caption{Experimental setup for CC-COP battery cycling.}
        \label{fig:exp_setup}
    \end{figure*}
    
    To extract degradation-related information, we perform the CC--COP control described in Section \ref{sec:CC-COP} and record the maximum terminal voltage $V_{\text{cut-off}}$ each time the battery reaches the target SoC$^*$. Data collection continues until the SoH of the battery falls below 80\%. This procedure can be carried out experimentally with the three-electrode PAT-Cell setup from ELCELL shown in Fig.~\ref{fig:exp_setup}, or replicated in a high-fidelity simulation. The resulting correlation between charge cut-off voltage and SoH provides important battery health information from the environment, and is used in the reward formulation for the battery charging agent. The core of our framework is a TD3-based agent that receives state and reward information from the environment and outputs optimal charging actions. Details of the agent formulation are presented in the following sections.

    In the implemented controller, the SoH--voltage mapping is stored as discrete pairs from the CC--COP trajectory and looked up by nearest-neighbour at the estimated SoH. The SoH--voltage mapping is cell-specific and requires calibration and validation on the target cell before practical use.

    \subsection{Overview of TD3 Method }\label{sec:RL_method}
    
    The lifelong charging task is sequential because each charging-current action affects the immediate SoC and voltage trajectory as well as the degradation state encountered in subsequent cycles. RL is used to learn this state-dependent, long-horizon trade-off offline, although it is not the only possible solution. TD3 is selected because the charging current is continuous and because its off-policy replay improves data reuse in computationally expensive electrochemical simulations. Its twin critics, delayed policy updates, and target-policy smoothing reduce value overestimation and policy-update error relative to DDPG \cite{fujimoto2018addressing}. These mechanisms improve empirical learning behaviour, but they do not constitute a formal proof of closed-loop stability, safety, or convergence for the nonlinear neural-network controller.
    
    Within the Actor-Critic framework, TD3 incorporates two types of networks: the critic network $Q(s,a|\theta^{Q})$ with parameter $\theta^{Q}$ and the actor network $\mu(s|\theta^{\mu})$ with parameter $\theta^{\mu}$. To mitigate the overestimation of the Q-value, TD3 employs a technique that calculates the target Q-value by selecting the smaller value from the two target critic networks, effectively reducing potential overestimation:
    \begin{equation}
        y_{t}(r, s^{\prime}) = r(s,a) +\gamma \underset{i=1,2}{\min} Q^{\prime}_{i}(s^{\prime}, a^{\prime}| \theta^{Q^{\prime}_{i}}),
    \end{equation}
    where $y_{t}$ is the target Q-value for given state $s$ and action $a$, $a^{\prime}$ is the next action chosen by the policy based on next state $s^{\prime}$, $\gamma$ is the discounting factor, and $\theta^{Q^{\prime}_{i}}$ is the parameter of target critic. The critic network updates its parameters based on gradient descent applied to the loss function as follows:
    \begin{equation}
        L(\theta^{Q_{i}}) = \mathbb{E} \left[ \left( y_{t} - Q_{i}\left(s, a | \theta^{Q_{i}} \right) \right)^2 \right],
    \end{equation}
    \begin{equation}
        \nabla_{\theta^{Q_{i}}} L(\theta^{Q_{i}}) = \mathbb{E} \left[ \left( y_{t} - Q_{i}(s, a | \theta^{Q_i}) \right) \nabla_{\theta^{Q_i}} Q_{i}(s, a | \theta^{Q_i}) \right],
    \end{equation}
    where \( \mathbb{E}(\cdot) \) denotes the expectation operator, and $\theta^{Q_{i}}$ is the parameter of $Q_{i}$. The parameter is updated (Update 1) according to:
    \begin{equation}
        \theta^{Q_i} \leftarrow \theta^{Q_i} - \alpha  \nabla_{\theta^{Q_{i}}} L(\theta^{Q_{i}}),
    \end{equation}
    where $\alpha$ is the learning rate of the critic network. Moreover, TD3 reduces instability by postponing updates to its actor network relative to its critic. In practice, the critic is updated frequently to accurately capture the dynamics of the environment, while the actor is adjusted only after a fixed number of critic updates. This delay minimises error propagation and keeps policy changes smoother, resulting in more stable and reliable learning outcomes. The actor network is updated (Update 2) with parameters using gradient ascent of the objective function as:
    \begin{equation}
        J_{Q}(\theta^\mu) = \mathbb{E}\left[ Q_1(s, \mu(s)) \right],
    \end{equation}
    \begin{equation}
        \nabla_{\theta^\mu} J_{Q}(\theta^\mu)
        = \mathbb{E}\left[ \nabla_a Q_1(s, a | \theta^{Q_1})\,\nabla_{\theta^\mu}\,\mu(s | \theta^\mu)\right],
    \end{equation}
    \begin{equation}
        \theta^\mu \leftarrow \theta^\mu \;+\; \beta\,\nabla_{\theta^\mu} J_{Q}(\theta^\mu),
    \end{equation}
    where $\theta^{\mu}$ is the parameter of actor network, $J(\theta^{\mu})$ is the defined objective function, $\nabla_{\theta^\mu} J(\theta^\mu)$ is the gradient of objective function with respect to $\theta^{\mu}$, $\beta$ is the learning rate of the policy network. In the soft target update (Update 3), the parameters of the target critic and the target actor network are updated as:
    \begin{gather}
        \theta^{Q^{\prime} }_{i} \leftarrow \tau \theta^{Q}_{i} + (1 - \tau) \theta^{Q^{\prime}}_{i},\\
        \theta^{\mu^{\prime}} \leftarrow \tau \theta^{\mu} + (1 - \tau) \theta^{\mu^{\prime}},
    \end{gather}
    where the soft updating factor is denoted by $\tau$ and $\theta^{\mu^{\prime}}$ is the parameter of the target actor network.
    
    Furthermore, TD3 introduces truncated normal noise to the actions produced by the target policy network, mitigating the trade-off between bias and variance. This added noise serves as a form of regularisation, which helps prevent overestimation of Q-values and reduces the risk of overfitting. By smoothing the target updates in this manner, TD3 achieves more stable learning and enhances the reliability of policy evaluation:
    \begin{equation}
        a_{t+1} \leftarrow \mu^{\prime}(s^{\prime}|\theta^{\mu^{\prime}}) + \varepsilon, \, \varepsilon \sim \text{clip} \left( \mathcal{N}(0, \sigma^2), a_{\min}, a_{\max} \right),
    \end{equation}
    where $a_{\min}$ and $a_{\max}$ define the valid range of actions, representing the lower and upper bounds of the battery charging current, respectively.
    
    \subsection{TD3 with Battery Implementation}\label{sec:reward_design}
    
    In the implemented TD3 controller, the state contains terminal voltage $V(t)$, the controller-observed SoC $\widetilde{\mathrm{SoC}}(t)$, and the filtered capacity-based SoH estimate $\widehat{\mathrm{SoH}}(t)$:
    \begin{equation}
        s(t)=[V(t),\widetilde{\mathrm{SoC}}(t),\widehat{\mathrm{SoH}}(t)]^{\top}.
    \end{equation}
    Before entering the actor, these three variables are linearly normalised from $[2.4,4.5]$~V, $[0,1]$, and $[0.5,1.1]$, respectively, to $[-1,1]$ and clipped to that interval. Here, $\widehat{\mathrm{SoH}}(t)$ is produced by the selected EMA or KF from the noisy SoH observation $\widetilde{\text{SoH}}(t)$, as described in Section~\ref{sec:soh_filtering}. The nominal simulations set $\widetilde{\mathrm{SoC}}(t)=\mathrm{SoC}(t)$, while the uncertainty tests perturb this observed quantity without exposing true SoC to the actor.

    These three state variables are selected for compatibility with signals available from a conventional BMS. SoC represents progress towards charge completion, terminal voltage reflects the instantaneous electrical boundary under load, and filtered SoH provides the ageing context used by the health-dependent reward.
    
    The action is the charging current, subject to the input constraint
    \begin{equation}
        a(t)=I(t),\qquad -10~\mathrm{A}\leq I(t)\leq0~\mathrm{A}.
    \end{equation}
    The actor output in $[-1,1]$ is linearly mapped to this interval. The negative sign denotes charging, and 10~A corresponds to 2C for the 5~Ah nominal cell.

    Safety is separated into hard electrical limits and soft degradation-aware reward shaping. The actor output is clipped to the admissible charging-current interval, and charging is terminated at the independent hard terminal-voltage limit. The SoH-dependent mapped voltage is a soft reward boundary rather than a certified plating-free constraint. Accordingly, the present TD3 policy does not guarantee the absence of lithium plating, and its safety evidence must be established empirically over the stated operating domain.

    Temperature is not included in the present state or reward because the evaluated environment assumes that a BMS-coordinated thermal-management system maintains the cell near a fixed reference temperature. This is a scope assumption rather than a claim that temperature is unimportant. Application under varying temperatures requires temperature to be included in the state, a temperature-dependent safety penalty or hard limit, a temperature-aware electrochemical and ageing environment, and retraining and validation over the intended thermal domain.

    The capacity-based SoH used here does not fully describe power capability or impedance growth. Terminal voltage provides indirect information about polarisation under load, but the controller does not explicitly use resistance, state of power, or heat generation. A power-aware extension should augment the state and safety limits with validated impedance and temperature information.
    
    The instantaneous reward is formulated as the sum of four components:
    \begin{equation} \label{eq:reward_all}
        r(t) = r_{\text{SoC}}(t) + r_{\text{volt}}(t) + r_{\text{SoH}}(t) + r_{\text{smooth}}(t),
    \end{equation}
    To encourage progress towards the target SoC, the first term is
    \begin{equation}
        r_{\text{SoC}}(t)=-w_{\mathrm{SoC}}\left|\mathrm{SoC}^{*}-\widetilde{\mathrm{SoC}}(t)\right|,
    \end{equation}
    where $\mathrm{SoC}^{*}$ is the target SoC and $w_{\mathrm{SoC}}$ is a positive weight.
    The mapped terminal-voltage value $V_{\mathrm{soft}}(\widehat{\mathrm{SoH}})$ is used as an SoH-dependent soft boundary:
    \begin{equation}
        r_{\text{volt}}(t)=-w_{\mathrm{volt}}\max\left[0,V(t)-V_{\mathrm{soft}}\bigl(\widehat{\mathrm{SoH}}(t)\bigr)\right],
    \end{equation}
    where $w_{\mathrm{volt}}$ is the voltage penalty weight.
    The controller obtains $V_{\mathrm{soft}}$ by nearest-neighbour lookup in the discrete SoH--voltage data and clamps an estimate outside the calibrated SoH range to the nearest endpoint. This value is a reward boundary, whereas the independent 4.5~V hard limit remains active. The SoH-related term combines an instantaneous current-magnitude surrogate with the observed long-term degradation trend:
    \begin{equation}
        r_{\text{SoH}}(t)=-w_{I}\left(\frac{|I(t)|}{I_{1\mathrm{C}}}\right)^{p}-w_{d}\,\rho_{\text{SoH}}(i),
    \end{equation}
    \begin{align}
        d_{\text{SoH}}(i)&=\frac{\widehat{\mathrm{SoH}}_{\mathrm{long}}(i-N+1)-\widehat{\mathrm{SoH}}_{\mathrm{long}}(i)}{N-1},\\
        \rho_{\text{SoH}}(i)&=\operatorname{clip}\left(\frac{d_{\text{SoH}}(i)}{d_{\mathrm{ref}}},\;\rho_{\min},\;\rho_{\max}\right),
    \end{align}
    where $w_{I}$ and $p$ are the weight and exponent for the current-magnitude penalty, and $w_{d}$ is the weight for the degradation trend. Here, $I_{1\mathrm{C}}$ is the nominal 1C-rate current, and $\widehat{\mathrm{SoH}}_{\mathrm{long}}$ is a dedicated long-term health estimate detailed in Section~\ref{sec:soh_filtering}. The first component discourages sustained high current without claiming to be a direct electrochemical degradation measurement. The second component compares the oldest and newest end-of-charge values in the method-specific buffer of $N$ completed charges and normalises their average decline by a reference per-cycle decline $d_{\mathrm{ref}}$. This reference value represents a standard 20\% capacity loss distributed over a nominal 600-cycle lifespan. The resulting ratio is then clipped to $[\rho_{\min},\rho_{\max}]$ to prevent a noisy trend estimate from dominating the reward. The archived EMA 5\% implementation uses the filtered SoH for the state and mapped voltage lookup but does not include the direct $r_{\text{SoH}}$ term. The EMA 5\% results correspond to this reward structure. Finally, the smoothness term is
    \begin{equation}
        r_{\text{smooth}}(t)=-w_{\mathrm{sm}}|I(t)-I(t-1)|,
    \end{equation}
    where $w_{\mathrm{sm}}$ is the smoothness penalty weight.
    The four dense terms have complementary roles: $r_{\mathrm{SoC}}$ drives charge completion, $r_{\mathrm{volt}}$ discourages operation above the health-dependent soft voltage boundary, $r_{\mathrm{SoH}}$ discourages high current and an adverse degradation trend, and $r_{\mathrm{smooth}}$ suppresses abrupt current changes. Their purpose is to express the competing charging, health, and control-regularity requirements, rather than to reconstruct an unmeasured plating state. These terms are examined individually in the reward ablation. As in the original implementation, a separate terminal constraint-shaping term is applied when a charge ends:
    \begin{equation}\label{eq:terminal_shaping}
        r_{\mathrm{done}}=
        \begin{cases}
        r_{\mathrm{term}}, & \text{target SoC reached},\\
        -(r_{\mathrm{term}}+w_{\Delta}\Delta_{\mathrm{SoC}}), & \text{premature termination},\\
        0, & \text{otherwise},
        \end{cases}
    \end{equation}
    where $r_{\mathrm{term}}$ is the positive base reward for successful completion, and $w_{\Delta}$ is the penalty slope for the SoC shortfall $\Delta_{\mathrm{SoC}}=\max(0,\mathrm{SoC}^{*}-\mathrm{SoC}_{\mathrm{true}})$. If the maximum number of control steps is reached without completion, the final-step reward is replaced by the timeout penalty $-r_{\mathrm{to}}$. This terminal shaping prevents early termination from becoming preferable to successful charging, but it does not override the validity rules. Any evaluation cycle that times out, terminates before reaching the true target SoC, or violates a hard constraint is still flagged as invalid and excluded from the reported statistics.

    \subsection{Robust SoH Processing via EMA and Kalman Filters}\label{sec:soh_filtering}
    
    In practical BMS deployments, the SoH estimate $\widetilde{\text{SoH}}(t)$ provided by onboard estimation algorithms is inevitably corrupted by error. The nominal simulation isolates relative zero-mean Gaussian SoH noise:
    \begin{align}
        \widetilde{\text{SoH}}(t)&=\text{SoH}(t)+\epsilon_{\text{SoH}}(t),\\
        \epsilon_{\text{SoH}}(t)&\sim\mathcal{N}\left(0,[\sigma_{\text{SoH}} \text{SoH}(t)]^2\right),\qquad \sigma_{\text{SoH}}\in\{0.02,0.05\}.
    \end{align}
    These levels are representative of the estimation-error range reported in \cite{fang2019online,shen2023accurate}. Directly using $\widetilde{\text{SoH}}(t)$ perturbs both the actor state and selection of the mapped soft voltage boundary. The 2\% and 5\% cases therefore test one practical source of map-selection error, although they do not quantify uncertainty in the offline calibration data.
    
    The EMA and KF are alternative implementations rather than two filters applied in series. EMA is used when reliable noise statistics are unavailable, whereas the KF is used when process and measurement covariances can be characterised. In either case, only the output of the selected filter is supplied to the state vector and reward computation.
    
    \subsubsection{Exponential Moving Average (EMA) Filter}
    
    The state EMA gives exponentially decreasing weights to older observations:
    \begin{equation}
        \widehat{\mathrm{SoH}}(t)=\alpha_{\mathrm{EMA}}\widetilde{\text{SoH}}(t)+(1-\alpha_{\mathrm{EMA}})\widehat{\mathrm{SoH}}(t-1).
    \end{equation}
    A second scalar EMA with $\alpha_{\mathrm{long}}=0.01$ produces $\widehat{\mathrm{SoH}}_{\mathrm{long}}$ for the 20-charge degradation-trend buffer, where $\alpha_{\mathrm{EMA}}=0.02$ for the state EMA. The two scalar recursions receive the same measurements but serve different purposes. A smaller coefficient produces stronger smoothing with greater lag.
    
    \subsubsection{Kalman Filter}
    
    Alternatively, the state KF models SoH as a random walk:
    \begin{align}
        \text{SoH}_k&=\text{SoH}_{k-1}+w_k,\qquad w_k\sim\mathcal{N}(0,Q),\\
        \widetilde{\text{SoH}}_k&=\text{SoH}_k+v_k,\qquad v_k\sim\mathcal{N}(0,R_k),
    \end{align}
    with $Q=(5\times10^{-6})^2$ and $R_k=[\sigma_{\text{SoH}}\max(\text{SoH}_k,0.5)]^2$ in the archived simulation code. For the additional frozen-policy evaluations, the measurement covariance is computed from observable quantities as $R_k=[\sigma_{\text{SoH}}\max(\widehat{\mathrm{SoH}}_{k|k-1},0.5)]^2$, where $\widehat{\mathrm{SoH}}_{k|k-1}$ is the prior filter estimate. The first noisy SoH measurement supplies the initial covariance reference. The covariance is initialised as $P_0=4R_0$ for each newly simulated cell. A second long-term KF supplies the reward-trend buffer using the same measurement equation and initialisation but $Q_{\mathrm{long}}=(4\times10^{-5})^2$. The 2\% and 5\% cases use the same process covariances, while $R_k$ changes with the imposed noise level.
    
    For either method, the state-filter output $\widehat{\mathrm{SoH}}(t)$ replaces the raw noisy observation in the actor state and the voltage-boundary lookup. Averaged over five training seeds and two evaluation-noise realisations per seed, the KF achieves an RMSE of $0.0062$ under 2\% noise and $0.0124$ under 5\% noise, a signed bias of $+0.0058$ and $+0.0119$ in SoH units, and zero cross-correlation lag in both cases. The steady-state gains at $\mathrm{SoH}=1$ are approximately $3.13\times10^{-4}$ for the 2\% case and $1.26\times10^{-4}$ for the 5\% case, meaning the filter places almost all weight on its internal prediction in steady state.

    The present uncertainty study perturbs only the SoH estimate and treats SoC as accurately available to isolate the effect of SoH filtering. It therefore does not establish robustness to joint SoC and SoH estimation errors. Deployment requires training and validation with representative SoC bias, drift, and noise characteristics from the target BMS estimator \cite{fang2019online}.

    \section{Results and Discussion}\label{sec:results}
    
    In this section, the proposed lifelong health-aware RL-based charging strategy is evaluated through data generated from high-fidelity simulations and benchmarked against multiple alternatives: the widely adopted CC--CV control, a health-aware CC-CV variant with SoH-linked voltage mapping (CC-CV-V), and baseline RL schemes with varying levels of SoH processing.
    
    \subsection{Battery Simulation Environment for Health-Aware Fast Charging}
    
    The proposed approach is model-free at deployment: no physically derived mathematical model is required as long as the state and reward signals can be obtained from the plant. However, conducting full-lifetime battery experiments to collect sufficient training data is highly time-consuming. We therefore employ a high-fidelity single-particle model with electrolyte (SPMe), using parameters from \cite{o2022lithium} and the open-source PyBaMM simulator \cite{sulzer2021python}, to generate data for training and evaluation. The partial differential-algebraic equations are discretised into a differential-algebraic equations via spectral methods \cite{li2021model, huang2024minn}. The underlying LG M50 electrochemical parameterisation was experimentally identified and evaluated against discharge curves at several currents \cite{chen2020parameterization}. The degradation submodels and associated parameter choices are adopted from \cite{o2022lithium}. These sources provide a documented and reproducible simulation configuration, but they do not constitute independent experimental validation of every ageing parameter or of the complete lifetime predictions reported here.
    
    \subsection{Experiment and Simulation Configurations}
    
    \begin{figure}[h!]
        \centering
        \begin{subfigure}[b]{0.75\linewidth}
            \centering
            \includegraphics[width=\linewidth,keepaspectratio]{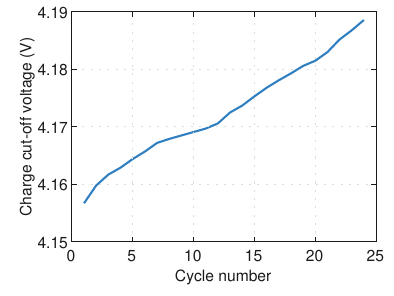}
            \caption{}\label{fig:vol_cutoff_exp}
        \end{subfigure}
        \hfill
        \begin{subfigure}[b]{0.75\linewidth}
            \centering
            \includegraphics[width=\linewidth,keepaspectratio]{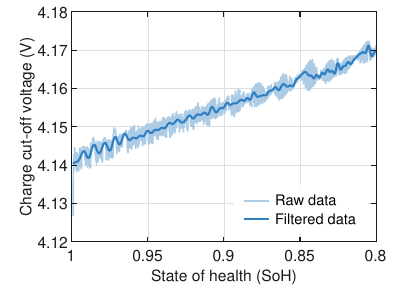}
            \caption{}\label{fig:voltage_end}
        \end{subfigure}
        \caption{Charge cut-off voltage against (a) cycle number in the three-electrode experiment and (b) SoH in the LG M50 simulation.}\label{fig:cut_off_SOH}
    \end{figure}

    \subsubsection{Mapping Acquisition between SoH and Charge Cut-off Voltage}
    
    To obtain the mapping between charge cut-off voltage and SoH, three-electrode NMC 811/graphite cells were tested using an EL-CELL PAT-Cell system, as illustrated in Fig.~\ref{fig:exp_setup}. The CC-COP charging process was conducted at 25~$^{\circ}$C in a constant-temperature chamber via a battery cycler, with an overpotential reference set to $\eta_\text{side}^* = 0.01$~V. During each cycle, the battery was charged to 80\% of its nominal capacity with the charging current regulated to keep the anode overpotential non-negative, and the cut-off voltage at charge termination was recorded. The experiment was repeated over multiple cycles, and the resulting charge cut-off voltage history is shown in Fig.~\ref{fig:vol_cutoff_exp}.
    
    Meanwhile, we applied the CC-COP method in a high-fidelity simulation to derive the mapping between SoH and cut-off voltage over the entire battery lifetime. The investigated battery is an LG M50 cell, with electrochemical parameters obtained from \cite{o2022lithium} and simulated using PyBaMM. The SPMe-ageing model is implemented, incorporating a single particle model with electrolyte dynamics to enable efficient and comprehensive analysis of cycle ageing. The primary degradation mechanisms are characterised by an SEI solvent diffusivity of $2.5 \times 10^{-22}$ m$^2$/s, a lithium plating kinetic rate constant $1 \times 10^{-11}$ m/s, and an irreversible condition for lithium plating.

    As shown in Fig.~\ref{fig:cut_off_SOH}, the experimental three-electrode cell and the LG M50 simulation both show a qualitative increase in the cut-off voltage with cell ageing. A similar trend under mild ageing was also reported in~\cite{epding2020aging}.
    
    \subsubsection{Controller Configuration and Training Results}\label{sec:controller_config}
    
    For the fast charging performance evaluation, we conduct simulations in the PyBaMM environment using an SPMe model configured with the published LG M50 electrochemical and degradation parameters \cite{chen2020parameterization,o2022lithium}. The learned policy and the mapping between SoH and voltage are specific to the chosen cell chemistry and operating domain. Transferring this approach to another chemistry requires target-specific parameter calibration, offline retraining, and independent experimental validation. A unified procedure is implemented to assess different charging strategies throughout the entire battery lifetime. The details of this procedure are outlined in Protocol~\ref{alg:charging_step}.

    \begin{algorithm}[tb]
        \caption{Battery cycling protocol for controller performance evaluation }\label{alg:charging_step}
        \begin{algorithmic}[1]
            \State \textbf{Initialise:} Discharge the battery to 0\% SoC.
            \State \textbf{Charge} at 1C until 4.2~V and hold at 4.2~V until C/20.
            \State \textbf{Discharge} at C/5 to 2.5~V to determine the initial capacity.
            \Repeat
            \State \textbf{Charge} from 0\% to $\mathrm{SoC}_0(i)=20\%$ at 2~A.
            \State \textbf{Apply} the target strategy immediately to charge from 20\% to $\mathrm{SoC}^{*}=80\%$.
            \State \textbf{Discharge} at 1C, reducing to 0.2C near the 2.5~V lower boundary for stable termination.
            \State \textbf{Rest} for one hour.
            \State \textbf{Check} SoH; if SoH falls to $\text{SoH}_{\text{end}}$ (e.g., 80\%), end test.
            \Until{battery SoH $\leq \text{SoH}_{\text{end}}$}
            \State \textbf{Evaluate} performance based on total cycles in terms of the averaged charging time and the equivalent full cycles.
        \end{algorithmic}
    \end{algorithm}
    
    To remain consistent with the published LG M50 degradation-model configuration, we adopt its ageing parameters and isothermal assumption \cite{o2022lithium}. We assume that a BMS-coordinated battery thermal management system maintains the cell temperature within a prescribed range around a fixed reference \cite{bao2025multiple,jang2025state}. Accordingly, all analyses and conclusions in this study are restricted to approximately isothermal operating conditions.
    
    In this study, we evaluate a set of charging strategies that systematically isolate the contributions of the proposed components. The conventional baselines are CC--CV and CC-CV-V, where the cut-off voltage of the latter is adapted via the SoH--voltage mapping. In the archived source used to generate the reported results, CC--CV applies a 9~A constant-current phase and a fixed $4.13$~V boundary, whereas CC-CV-V applies an 8.5~A constant-current phase and dynamically adjusts the voltage boundary using the SoH--voltage mapping. In addition, two advanced health-aware benchmarks are included: a multistage constant-current (Multi-CC) strategy and a model predictive control (MPC) baseline. Multi-CC implements a three-stage current profile: 9.0~A (1.8C) for 20\%--40\% SoC, 7.0~A (1.4C) for 40\%--60\% SoC, and 4.0~A (0.8C) for 60\%--80\% SoC. The MPC baseline utilises an equivalent-circuit OCV--resistance model over the 20~s control interval to analytically solve for the maximum allowable charging current satisfying the health-dependent voltage limit $V_{\mathrm{soft}}(\mathrm{SoH})$ and physical limits. Its fixed model parameters are a resistance of 0.060~$\Omega$ and an OCV slope of 0.50~V per unit SoC, with a 9~A current limit. Degradation awareness enters through the CC--COP-derived SoH--voltage constraint, without anode-overpotential feedback. To isolate the effect of the SoH-aware reward terms, we include a partial-RL baseline under 2\% SoH noise that uses only $r_{\text{SoC}}$ and $r_{\text{volt}}$ with unfiltered SoH, and its EMA-filtered counterpart that applies the same reward structure.

    The actor and each critic contain two fully connected hidden layers with 200 ReLU units per layer. The evaluations use the original method-specific implementations. They include the terminal shaping term in (\ref{eq:terminal_shaping}). The four-term reward ablation and coefficient-sensitivity experiments use the archived KF 2\% implementation as their fixed reference core and alter only the named reward term or coefficient. The external control interval is 20~s. The SPMe spatial discretisation uses five points through each negative-electrode, separator, and positive-electrode domain and ten radial points in each particle. After each target charge, discharge uses 5~A with 180~s solver steps above 3.0~V, 5~A with 30~s steps from 3.0 to 2.8~V, and 1~A with 10~s steps below 2.8~V until 2.51~V, followed by a 1~h rest. Each newly trained policy is trained for 2000 charge episodes. A training cell is replaced when its true SoH falls below 70\%, whereas reported lifetime evaluation stops at 80\% SoH and contains no exploration noise. Training exploration combines an epsilon-random action, with $\epsilon_i=\max(0.15\times0.997^i,0.02)$, and Gaussian action noise whose standard deviation follows $\max(0.25\times0.999^i,0.10)$ in the normalised action space. The complete TD3 settings are listed in Table~\ref{tab:TD3_parameters}.

    Because the RL policy is trained offline, only the trained actor network and the selected scalar filter are required online. The actor is a lightweight feed-forward neural network comprising two hidden layers with 200 units each. The critics, target networks, optimiser states, and replay buffer are required only during offline training and are not deployed. This feed-forward architecture imposes a modest computational footprint relative to the 20~s control interval. However, direct embedded deployability, including exact inference latency and peak memory usage, requires future experimental evaluation on target BMS hardware.

    The reward coefficients introduced in Section~\ref{sec:reward_design} are set as follows. For the dense terms: $\mathrm{SoC}^{*}=0.8$, $w_{\mathrm{SoC}}=2$, $w_{\mathrm{volt}}=5$, $w_{I}=0.55$, $p=1.5$, $I_{1\mathrm{C}}=5$~A, $w_{d}=0.05$, $d_{\mathrm{ref}}=0.2/600$, $\rho_{\min}=-1$, $\rho_{\max}=3$, and $w_{\mathrm{sm}}=0.15$. For the terminal shaping: $r_{\mathrm{term}}=20$, $w_{\Delta}=200$, and $r_{\mathrm{to}}=30$. The degradation-trend buffer lengths are $N=20$ (EMA 2\%), $N=15$ (KF 2\%), and $N=30$ (KF 5\%). These values were determined through preliminary training experiments to balance the competing reward components and were held fixed for all subsequent evaluations.

    \begin{table}[tb]
        \caption{Hyperparameters of the Proposed TD3 Algorithm}\label{tab:TD3_parameters}
        
        \begin{centering}
            \begin{tabular}{cc}
                \hline \hline
                Hyperparameter & Values\tabularnewline
                \hline
                Actor network learning rate & 0.0001\tabularnewline
                Critic network learning rate & 0.0001\tabularnewline
                Discounting factor & 0.995\tabularnewline
                Experience replay buffer size & $1.0 \times 10^{6}$ \tabularnewline
                Mini-batch size & 256\tabularnewline
                Soft update factor & 0.005\tabularnewline
                Delay frequency & 2\tabularnewline
                Target-policy noise / clip & 0.1 / 0.1\tabularnewline
                Hidden layers / units & 2 / 200\tabularnewline
                Training episodes & 2000\tabularnewline
                Activation function & ReLU\tabularnewline
                Optimiser & Adam\tabularnewline
                \hline \hline
            \end{tabular}
            \par\end{centering}
    \end{table}

    \begin{figure}[htb]
        \centering
        \begin{subfigure}[b]{0.75\linewidth}
            \centering
            \includegraphics[width=\linewidth,keepaspectratio]{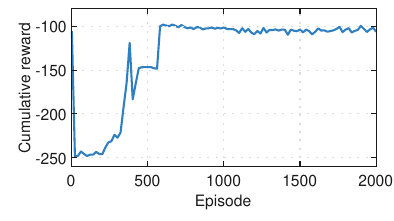}
            \caption{}\label{fig:rl_reward_kf_5soh}
        \end{subfigure}
        \hfill
        \begin{subfigure}[b]{0.75\linewidth}
            \centering
            \includegraphics[width=\linewidth,keepaspectratio]{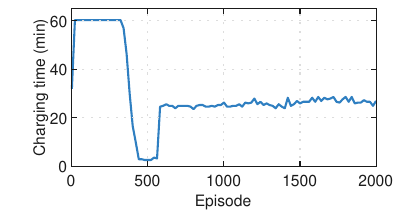}
            \caption{}\label{fig:rl_time_charge_kf_5soh}
        \end{subfigure}
        \caption{Training performance of the proposed KF variant under 5\% SoH estimation noise: (a) cumulative rewards; (b) charging time.}\label{fig:RL_training}
    \end{figure}
    
    The training performance of the proposed charging algorithm is presented in Fig.~\ref{fig:RL_training}, illustrating the evolution of two metrics across 2000 training episodes. Fig.~\ref{fig:rl_reward_kf_5soh} shows the cumulative reward, which begins at approximately $-250$ and improves substantially between episodes 300 and 600, with transient oscillations during this learning phase. The reward then stabilises around $-100$ from episode 600 onward, indicating convergence of the policy. Fig.~\ref{fig:rl_time_charge_kf_5soh} shows the corresponding charging time evolution. Initially capped at 60~min due to the untrained policy, the charging time exhibits a brief transient dip to near zero around episode 500, reflecting an intermediate exploratory phase, before rising and stabilising at approximately 25~min from episode 600 onward. The concurrent convergence of both metrics confirms that the agent has successfully learned to balance charging speed against the health and safety penalties in the reward. 

    \subsection{Fast Charging Results and Overall Analysis}
    
    \begin{figure}[tb]
        \centering
        \includegraphics[width=0.8\linewidth,keepaspectratio]{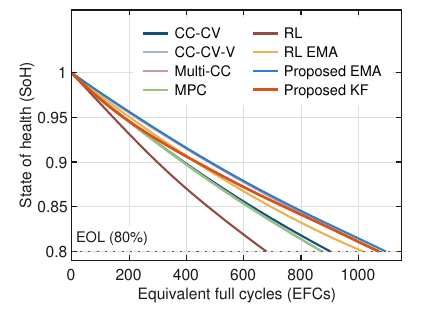}
        \caption{Comparison of SoH degradation versus equivalent full cycles (EFCs) among different charging methods under 2\% SoH estimation noise.}
        \label{fig:SoH_vs_EFCs}
    \end{figure}
    
    Following the battery cycling protocol as described in Protocol~\ref{alg:charging_step}, we conducted the battery charging tests using all strategies defined in Section~\ref{sec:controller_config}. To balance fidelity and efficiency, the sampling interval for all lifelong implementations is chosen at 20\,s. EFCs and average charging duration serve as the primary performance metrics, reflecting battery lifetime and charging speed, respectively. For charging-time statistics, the first five cycles are excluded before applying a three-interquartile-range outlier filter and calculating the mean.
    
    \subsubsection{Overall Analysis on Charging Time and Lifespan Under 2\% Noise}
    
    \begin{table}[tb]
        \centering
        
        \caption{Battery Charging Performance Comparison Under 2\% SoH Estimation Noise}
        \label{tab:EFCs_time}
        \setlength{\tabcolsep}{5pt}
        \renewcommand{\arraystretch}{1.15}
        \begin{tabular}{lccc}
            \hline\hline
            Method & \makecell{Avg.\ time\\(min)} & EFCs & \makecell{Rel.\ EFCs\\vs.\ CC--CV} \\
            \midrule
            CC--CV        & 26.20 & 902.21 & Ref. \\
            CC-CV-V       & 24.47 & 872.98 & $-3.2\%$ \\
            Multi-CC      & 26.86 & 1074.23 & $+19.1\%$ \\
            MPC           & 23.06 & 875.11 & $-3.0\%$ \\
            RL            & 24.88 & 679.14 & $-24.7\%$ \\
            RL EMA        & 24.79 & 1017.05 & $+12.7\%$ \\
            Proposed EMA  & 25.97 & 1095.66 & $+21.4\%$ \\
            Proposed KF   & 26.03 & 1071.34 & $+18.7\%$ \\
            \hline\hline
        \end{tabular}
    \end{table}
    
    We first analyse the performance of all strategies under 2\% SoH estimation noise, with quantitative results in Table~\ref{tab:EFCs_time} and degradation trajectories in Fig.~\ref{fig:SoH_vs_EFCs}. Three performance tiers emerge: the rule-based strategies establish the reference; the baseline RL results isolate the effect of SoH signal quality; and the proposed variants demonstrate the combined benefit of the full health-aware reward and robust filtering.
    
    CC-CV establishes the conventional baseline, applying a fixed 9~A current with a constant cut-off voltage throughout battery life. CC-CV-V dynamically adapts the cut-off voltage according to the SoH--voltage mapping. Because both strategies share the same SoC target, they deliver nearly identical charge per cycle. The shorter charging time of CC-CV-V is instead driven by the adaptive voltage limit, which sustains a higher average current and shortens the constant-voltage taper. However, this more aggressive current profile accelerates cell degradation, causing CC-CV-V to reach end of life earlier (872.98 EFCs) than the standard CC-CV baseline (902.21 EFCs). This demonstrates that adapting the voltage limit alone, without jointly optimising the current profile, incurs a higher degradation cost per cycle and provides no net lifetime benefit.
    
    The baseline RL strategy, which ingests noisy SoH directly without filtering, exhibits the most severe degradation among all evaluated strategies, achieving roughly a quarter fewer EFCs than CC-CV. This is likely attributable to the noisy SoH corrupting the reward signal during training, undermining the consistency of policy updates. Critically, this result occurs despite the RL architecture being otherwise identical to the proposed method, which suggests that SoH signal quality is the decisive factor rather than the learning algorithm itself. Applying EMA filtering to the same reward structure recovers a substantial portion of the lost performance, confirming that SoH signal conditioning plays a critical role in stable, health-aware RL control.
    
    Among the time-comparable strategies in Table~\ref{tab:EFCs_time}, the proposed EMA achieves the highest lifetime at 1095.66 EFCs and 25.97~min. Relative to Multi-CC, this gives 2.0\% more EFCs with a 3.3\% shorter charging time. The proposed KF attains a lifetime within 0.3\% of Multi-CC while charging 3.1\% faster. The consequence is a consistently slower SoH degradation trajectory visible throughout the entire service life, as shown in Fig.~\ref{fig:SoH_vs_EFCs}. The proposed EMA and KF variants achieve 21.4\% and 18.7\% more EFCs than CC-CV respectively, while maintaining comparable charging times, indicating that the SoH-aware reward terms and robust filtering each contribute meaningfully to the overall performance gain.

    \subsubsection{Degradation and Safety Indicators}\label{sec:degradation_indicators}
    To further evaluate the safety boundaries and the underlying mechanisms of lifetime extension, we analyse three direct degradation indicators: the minimum anode overpotential, the cumulative capacity loss due to lithium plating, and the cumulative capacity loss due to solid electrolyte interphase (SEI) growth. These metrics provide a more detailed understanding of the ageing process than the terminal EFCs alone.

\begin{figure}[tb]
        \centering
        \begin{subfigure}[b]{0.75\linewidth}
            \centering
            \includegraphics[width=\linewidth,keepaspectratio]{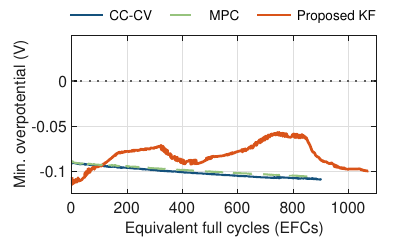}
            \caption{}
        \end{subfigure}
        
        \vspace{0.2em}
        \begin{subfigure}[b]{0.75\linewidth}
            \centering
            \includegraphics[width=\linewidth,keepaspectratio]{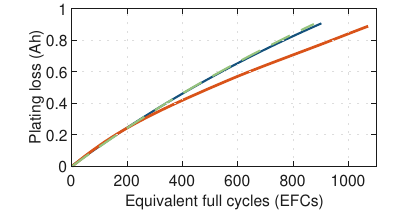}
            \caption{}
        \end{subfigure}
        
        \vspace{0.2em}
        \begin{subfigure}[b]{0.75\linewidth}
            \centering
            \includegraphics[width=\linewidth,keepaspectratio]{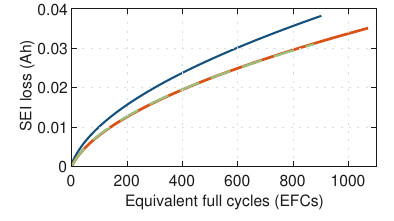}
            \caption{}
        \end{subfigure}
        \caption{Evolution of degradation indicators across the battery lifespan: (a) minimum anode overpotential, (b) cumulative lithium plating capacity loss, and (c) cumulative SEI capacity loss.}
        \label{fig:degradation}
    \end{figure}

    Fig.~\ref{fig:degradation} compares the evolution of these indicators across the battery lifespan for the CC-CV baseline, the MPC method, and the proposed KF strategy. As shown in Fig.~\ref{fig:degradation}(a), the proposed KF strategy exhibits a slightly lower minimum anode overpotential during the early life cycles, indicating a more aggressive initial charging profile. However, as the battery ages, the proposed strategy dynamically elevates the minimum anode overpotential, achieving a lifetime-averaged minimum of $-0.082$~V compared to $-0.102$~V for CC-CV and $-0.099$~V for MPC. The higher lifetime-averaged minimum anode overpotential is consistent with reduced plating exposure under the tested conditions.

    The benefits of this adaptive behaviour are reflected in the cumulative degradation losses. As shown in Fig.~\ref{fig:degradation}(b), the proposed KF accumulates 0.889~Ah of plating-related capacity loss at end of life, compared with 0.907~Ah for CC-CV and 0.903~Ah for MPC, while sustaining 18.7\% more EFCs than CC-CV. In Fig.~\ref{fig:degradation}(c), its SEI-related capacity loss is 0.035~Ah, compared with 0.038~Ah for CC-CV and 0.031~Ah for MPC. These indicators support lifetime extension with reduced plating-related capacity loss relative to CC-CV in the evaluated cell model.

    \subsubsection{Reward-Term Ablation}

    \begin{table}[tb]
        \centering
        \caption{Reward ablation under 2\% SoH estimation noise.}
        \label{tab:reward_ablation}
        \setlength{\tabcolsep}{3pt}
        \renewcommand{\arraystretch}{1.15}
        \scriptsize
        
        \resizebox{\columnwidth}{!}{\begin{tabular}{lccc}
            \hline\hline
            \multicolumn{4}{c}{\textit{(a) Charging performance}}\\
            Reward configuration & Valid runs & EFCs & Avg. time (min) \\
            \midrule
            Full reward                    & 3/3 & $1036.8\pm96.4$  & $26.90\pm4.43$ \\
            Without $r_{\mathrm{SoC}}$      & 0/3 & Infeasible       & -- \\
            Without $r_{\mathrm{volt}}$     & 3/3 & $1010.5\pm166.2$ & $26.45\pm5.83$ \\
            Without $r_{\mathrm{SoH}}$      & 3/3 & $912.8\pm49.7$   & $22.32\pm1.92$ \\
            Without $r_{\mathrm{smooth}}$   & 3/3 & $1137.7\pm93.1$  & $28.26\pm1.02$ \\
            \hline\hline
        \end{tabular}}

        \vspace{0.5em}

        \resizebox{\columnwidth}{!}{\begin{tabular}{lccc}
            \hline\hline
            \multicolumn{4}{c}{\textit{(b) Electrical and control metrics}}\\
            Reward configuration & $V_{\max}$ (V) & \makecell{Min. anode\\overpotential (V)} & $\Sigma|I_k-I_{k-1}|$ (A) \\
            \midrule
            Full reward                    & $4.2609\pm0.0668$ & $-0.1195\pm0.0135$ & $9.62\pm1.34$ \\
            Without $r_{\mathrm{SoC}}$      & --                & --                  & -- \\
            Without $r_{\mathrm{volt}}$     & $4.2934\pm0.0350$ & $-0.1388\pm0.0284$ & $11.75\pm4.28$ \\
            Without $r_{\mathrm{SoH}}$      & $4.2597\pm0.0873$ & $-0.1208\pm0.0146$ & $7.03\pm0.95$ \\
            Without $r_{\mathrm{smooth}}$   & $4.2117\pm0.0244$ & $-0.0985\pm0.0261$ & $13.93\pm8.14$ \\
            \hline\hline
        \end{tabular}}

        {\raggedright\footnotesize\textsuperscript{*}Each entry is the mean $\pm$ sample standard deviation over three independent training runs.\par}
    \end{table}

    Table~\ref{tab:reward_ablation} reports a one-term-at-a-time ablation using the fixed KF 2\% core and 2000 training episodes. Each configuration contains three independent training runs and one evaluation realisation per trained policy. The source, hyperparameters, and evaluation protocol are unchanged, and only the named reward coefficient is set to zero. Statistics are calculated across the three trained policies. Given the number of independent runs, the results are interpreted as descriptive ablation evidence rather than a formal significance test.

    A policy is valid only when the target SoC and all hard constraints are satisfied. Without $r_{\mathrm{SoC}}$, all three policies fail to reach the target SoC in every evaluation cycle. This outcome is structural because $r_{\mathrm{SoC}}$ is the only term that drives SoC advancement, and removing it eliminates the agent's incentive to complete the charge. Removing $r_{\mathrm{SoH}}$ reduces mean EFCs by 12.0\% while shortening mean charging time by 17.0\%, which identifies its role in the lifetime--speed trade-off. Removing $r_{\mathrm{volt}}$ changes mean EFCs by $-2.5\%$, but increases the EFC standard deviation from 96.4 to 166.2, raises the maximum terminal voltage from 4.2609 to 4.2934~V, and shifts the minimum anode overpotential from $-0.1195$ to $-0.1388$~V. These changes are consistent with the intended safety-oriented shaping role of the voltage term. Removing $r_{\mathrm{smooth}}$ increases mean EFCs by 9.7\%, but lengthens charging by 5.1\% and increases the mean per-cycle current total variation by 44.8\%. Thus, the smoothness term regularises the control trajectory rather than directly maximising lifetime. The complete reward is not optimal for every individual metric, but provides a feasible compromise among completion, charging speed, lifetime, voltage-related risk, and current regularity.

    \subsubsection{Impact of SoH Noise Level on Proposed Methods}
    
    \begin{figure}[tb]
        \centering
        \includegraphics[width=0.8\linewidth,keepaspectratio]{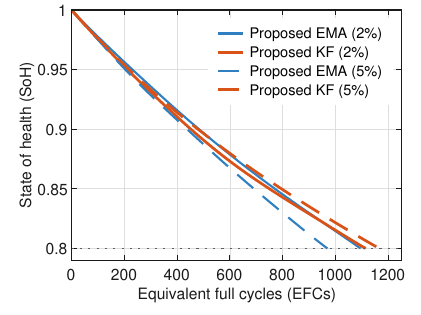}
        \caption{Representative single-policy SoH trajectories under 2\% and 5\% SoH noise. The same colour indicates the same method. Solid lines indicate 2\% noise and dashed lines indicate 5\% noise.}
        \label{fig:SoH_vs_EFCs_noise}
    \end{figure}
    
    To evaluate the robustness of the proposed methods against varying levels of SoH estimation noise, we compare the proposed EMA and KF variants under both 2\% and 5\% noise conditions, with results shown in Fig.~\ref{fig:SoH_vs_EFCs_noise}.
    
    Under 2\% SoH noise, both proposed variants extend lifetime, with EMA achieving 1095.66 EFCs and KF achieving 1071.34 EFCs in Table~\ref{tab:EFCs_time}. The KF achieves a longer lifetime than EMA under 5\% noise in the representative trajectories of Fig.~\ref{fig:SoH_vs_EFCs_noise}. This supports using EMA when noise statistics are unavailable and KF when they can be characterised.
    
    In the representative 5\% noise run, the KF variant achieves 1166.77 EFCs, 29.3\% more than CC--CV, at an average charging time of 26.22~min. The repeated-policy analysis below evaluates training variability separately from this single-policy illustration.

    To assess robustness to training variability, the evaluation was repeated over five independent training seeds with two evaluation-noise realisations per seed. The across-seed statistics are reported in Table~\ref{tab:kf_uncertainty}. The mean KF lifetimes exceed 1000 EFCs at both noise levels. Using each trained policy's mean over its two evaluations as one independent observation, a two-sided Welch test gives $p=0.0557$ for the 5\%-minus-2\% mean difference of 110.9 EFCs, with a 95\% confidence interval from $-3.6$ to 225.5 EFCs. No difference is detected at the 0.05 significance level.

    \begin{table}[tb]
        \centering
        \caption{Multi-seed uncertainty analysis for the proposed KF under 2\% and 5\% SoH noise.}
        \label{tab:kf_uncertainty}
        \setlength{\tabcolsep}{5pt}
        \renewcommand{\arraystretch}{1.15}
        
        \begin{tabular}{lcc}
            \hline\hline
            \makecell[l]{SoH noise\\level} & EFCs & \makecell{Avg.\ charging\\time (min)} \\
            \midrule
            2\% & $1018.1 \pm 91.5$ & $26.4 \pm 3.7$ \\
            5\% & $1129.0 \pm 55.2$ & $27.4 \pm 2.5$ \\
            \hline\hline
        \end{tabular}
        
    \end{table}

    Additional evaluations freeze the five trained KF policies at each noise level and use the observable covariance in Section~\ref{sec:soh_filtering}. With independent Gaussian errors, mean lifetimes are 1017.7 and 1128.8 EFCs under 2\% and 5\% noise. When a 1\% positive relative bias, AR(1) correlation of 0.9 and once-per-cycle SoH updates are applied together, the corresponding means remain 1003.8 and 1117.3 EFCs. All evaluation cycles reach the target SoC. These results confirm the lifetime benefit under the tested estimator-error conditions without retraining the actor.

    Under 5\% noise, the representative EMA run reaches 969.87 EFCs, compared with 902.21 EFCs for CC--CV and 1166.77 EFCs for KF. Thus, the KF achieves a larger lifetime gain under 5\% estimation noise, while EMA remains a viable option when reliable noise statistics are unavailable.
    
    \begin{figure}[tb]
        \centering
        \includegraphics[width=0.8\linewidth,keepaspectratio]{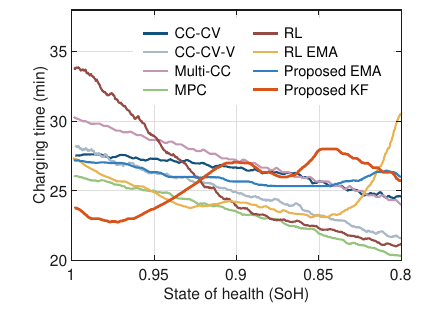}
        \caption{Charging time evolution as a function of SoH for all charging methods.}
        \label{fig:time_vs_SoH}
    \end{figure}
    
    \subsubsection{Charging Time Dynamics Over Battery Ageing}
    
    Fig.~\ref{fig:time_vs_SoH} illustrates how the charging time of each method evolves as the battery ages. Traditional fixed-rule strategies, including CC--CV and Multi-CC, exhibit a continuous decrease in charging time over the battery lifespan. This is not an adaptive feature but a byproduct of capacity fade and resistance growth, which cause the cell to reach voltage limits earlier and terminate the charging process sooner. The MPC baseline, despite its internal model, also follows this natural monotonic decrease. 
    
    In contrast, the proposed KF strategy exhibits a fundamentally different, actively adaptive profile: it charges aggressively early in life (resulting in shorter charging sessions) and becomes increasingly conservative as the battery ages, intentionally prolonging charging time in the later stages (e.g., between 90\% and 80\% SoH). This emergent behaviour is consistent with the degradation analysis in Section~\ref{sec:degradation_indicators}. By exploiting the higher tolerance of a healthy cell to accept fast charging, the agent minimises charging time early in battery life. As the cell ages and becomes more vulnerable to lithium plating and internal stresses, the policy shifts its priority towards health conservation, actively slowing down the charge to protect the degraded cell. This age-dependent trade-off between charging speed and lifetime demonstrates the adaptation learned by the SoH-aware RL formulation, arising naturally from the long-horizon objective without explicit rule-based programming.
    
    \subsubsection{Evolution of Current and Overpotential Profiles}
    
    \begin{figure}[tb]
        \centering
        \begin{subfigure}[b]{0.75\linewidth}
            \centering
            \includegraphics[width=\linewidth,keepaspectratio]{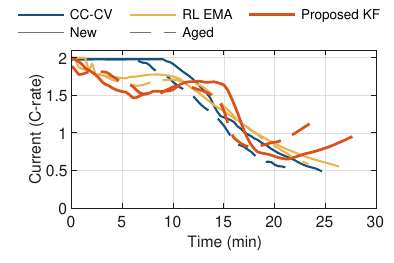}
            \caption{}
            \label{fig:CC-current_cycle}
        \end{subfigure}
        \hfill
        \begin{subfigure}[b]{0.75\linewidth}
            \centering
            \includegraphics[width=\linewidth,keepaspectratio]{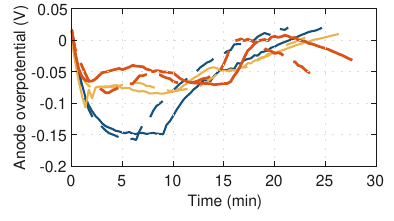}
            \caption{}
            \label{fig:eta_cycle}
        \end{subfigure}
        
        \caption{Current and anode overpotential profiles of the studied cell at varying ageing levels under different control methods.}
        \label{fig:result_cycles}
    \end{figure}

    The profiles in Fig.~\ref{fig:result_cycles} provide further insight into the behaviour of each strategy. The proposed RL method produces a multi-stage current pattern that resembles a stepped CC-COP profile but is not explicitly prescribed by the reward. The proposed KF reduces the depth of negative anode overpotential relative to CC-CV at both illustrated ageing stages. The anode overpotential is shown here for analytical purposes only and is not used as a control input.
    
    A notable observation is the similarity between the current profiles for new and aged cells under the proposed method, which is less apparent in CC-CV. For CC-CV, as internal resistance rises with ageing, the anode overpotential in aged cells spends more time at deeply negative values compared to a fresh cell, which is associated with greater lithium plating risk. The proposed method, by conditioning on real-time SoH, produces current profiles that remain more consistent across ageing stages, as visible in Fig.~\ref{fig:result_cycles}.
    
    As noted in Section~\ref{sec:problem_charging}, this work does not impose strict limits on anode overpotential, acknowledging that some plating is unavoidable at fast-charging rates. The relevant question is whether a strategy exacerbates or reduces its occurrence relative to the baseline. The overpotential profiles in Fig.~\ref{fig:result_cycles} suggest the proposed approach is more conservative in this regard.

    \section{Conclusion}\label{sec:conclusion}

    This work demonstrates filter-augmented SoH feedback for lifelong health-aware fast charging. An RL agent trained on unfiltered, noisy SoH performs worse than conventional CC--CV, while the same architecture with EMA filtering achieves a 21.4\% EFC gain and with KF filtering achieves 18.7\% under 2\% estimation noise, both at comparable charging speeds. Under 5\% estimation noise, the KF maintains the lifetime advantage and achieves higher EFCs than EMA in the representative comparison. The observable-covariance evaluations confirm the lifetime benefit with biased, correlated and intermittently updated SoH feedback. These results demonstrate the value of filtered SoH feedback for lifelong charging, and the KF should be preferred when noise statistics can be characterised. EMA is an adequate, computationally simpler option for well-calibrated BMS estimators where noise statistics are unavailable.
    
    Beyond the filtering result, the RL policy exhibits an emergent health-investment behaviour: the agent charges aggressively early in battery life and becomes increasingly conservative as the cell ages, a strategy arising without explicit programming from the long-horizon objective. This learned adaptation balances charging speed and lifetime across battery ageing. The experimental SoH--voltage trend was obtained from a three-electrode NMC 811/graphite cell, which differs from the simulated LG M50 cell. Both cells show a rise in the charge cut-off voltage with ageing. For practical deployment, the target-cell mapping requires calibration and validation through the CC--COP procedure before the framework is applied. Its online computation is limited to actor-network inference and scalar-filter updates. The architecture is compatible with resource-constrained implementation in principle, but direct embedded deployability requires target-hardware measurements of inference latency, memory use, and numerical behaviour.

    The conclusions of this study are limited to the assumed thermal-management scenario coordinated by the BMS, in which the cell temperature remains within a prescribed range around a fixed reference. Extending the framework to varying temperatures, through a temperature-aware model, controller retraining, and independent validation, is left as future work.

    \vfill

\end{document}